\documentclass[runningheads]{llncs}
\usepackage[T1]{fontenc}
\usepackage{graphicx}
\usepackage{booktabs}
\usepackage{array}
\usepackage{multirow}
\begin{document}
\title{Multi-Sample Prompting and Actor-Critic Prompt Optimization for Diverse Synthetic Data Generation}
\titlerunning{Multi-Sample Prompting and Actor-Critic Optimization for Diverse SDG}
\author{Abdelkarim El-Hajjami\orcidID{0009-0004-7053-3264} \and \\
Camille Salinesi\orcidID{0000-0002-1957-0519}}
\authorrunning{A. El-Hajjami and C. Salinesi}
\institute{Paris 1 Panth\'{e}on--Sorbonne University, Paris, France\\
\email{\{abdelkarim.el-hajjami, camille.salinesi\}@univ-paris1.fr}}
\maketitle
\begin{abstract}
High-quality labeled datasets are fundamental for training and evaluating machine learning models, yet domains such as healthcare and Requirements Engineering (RE) face persistent barriers due to data scarcity, privacy constraints, or proprietary restrictions. While Large Language Models (LLMs) offer a promising avenue for Synthetic Data Generation (SDG), LLM-generated data tends to be repetitive and low in diversity, reducing its effectiveness for downstream tasks. Two approaches show potential for addressing this limitation: (1) multi-sample prompting, which generates multiple samples per prompt to reduce repetition, and (2) Prompt with Actor-Critic Editing (PACE), which iteratively refines prompts to maximize diversity. We integrate both mechanisms into Synthline, a Feature Model-based configurable synthetic data generator, and assess their effects on diversity and downstream utility across four RE classification tasks. Multi-sample prompting consistently improves both diversity and utility, with F1-score gains of 6 to 43.8 percentage points. PACE-based prompt optimization consistently improves lexical diversity but produces task-dependent utility effects, revealing the risks of optimizing for diversity alone. Most notably, synthetic data can match or surpass human-authored data for tasks where real labeled data is limited, with improvements of up to 15.4 percentage points in F1-score.
\end{abstract}
\section{Introduction}
\label{sec:intro}
High-quality labeled datasets are fundamental for training and evaluating machine learning models, yet acquiring such data remains costly, time-consuming, and often restricted by privacy or proprietary constraints~\cite{ref_fs}. Large Language Models (LLMs) have emerged as a practical tool for Synthetic Data Generation (SDG), offering the ability to produce labeled training examples on demand~\cite{ref_surveyllm}.

LLM-based SDG, however, faces a core limitation: the generated data tends to be repetitive and low in diversity, reducing its effectiveness for downstream tasks~\cite{ref_fs,zhang2025verbalized}. Early SDG approaches relied on zero-shot prompting~\cite{ref_zs}, where an LLM generates task-specific data without examples. Few-shot prompting~\cite{ref_fs} improved control by conditioning generation on a small set of labeled samples. Attributed prompting~\cite{ref_attr} improves diversity over simple class-conditional prompting by specifying data attributes (e.g., domain, style, length) in the prompt. Our prior work on Synthline~\cite{ref_synthlinev0}, a Feature Model (FM)-based configurable synthetic data generator, systematized this approach by enabling fine-grained control over generation attributes. However, despite this, the generated data still exhibited lower diversity than real data, indicating that attribute-based approaches alone are insufficient to address the diversity limitation. Zhang et al.~\cite{zhang2025verbalized} trace this limitation to post-training alignment, which introduces typicality bias: human annotators favor familiar text, causing LLMs to converge toward stereotypical outputs and leading to mode collapse. The problem we address in this paper is: \textit{how can we improve the diversity of LLM-generated synthetic data while preserving or improving its downstream utility for classification tasks?}

Two approaches show potential for addressing this limitation. First, multi-sample prompting, requesting multiple data points within a single prompt, has been shown to improve diversity in paraphrase generation~\cite{berro2025llms}. Second, automatic prompt optimization techniques such as Prompt with Actor-Critic Editing (PACE)~\cite{ref_pace} offer a promising direction for SDG~\cite{freise2025automatic}. Neither approach has been evaluated for SDG.

In this paper, we address this gap. We integrate multi-sample prompting and PACE into Synthline, and evaluate their effects on both the diversity and the downstream classification utility of the generated data. We formulate the following research questions:
\begin{itemize}
    \item \textbf{RQ1:} How does multi-sample prompting impact the diversity and utility of synthetic data for classification tasks?
    \item \textbf{RQ2:} How does PACE-based prompt optimization influence the diversity and utility of synthetic data for classification tasks?
    \item \textbf{RQ3:} To what extent can synthetic data replace or surpass human-authored data for classification tasks?
\end{itemize}

We evaluate both \textit{diversity} and \textit{utility} because they capture complementary quality dimensions of synthetic training data~\cite{ref_fs,ref_surveyllm}. Diversity measures whether the LLM produces varied outputs---necessary because repetitive synthetic data underrepresents the variability of real-world instances, which can limit the effectiveness of classifiers trained on it~\cite{ref_fs,zhang2025verbalized}. Utility measures whether the generated data actually improves downstream task performance, which is the ultimate goal of SDG. Evaluating both is essential because they do not always align: increasing diversity through techniques such as temperature sampling can degrade label accuracy~\cite{ref_diversityaccuracy}, while low-diversity data may overfit to narrow patterns.

We ground our study in the domain of Requirements Engineering (RE), where data scarcity is a well-documented challenge. The shortage of publicly available, labeled datasets remains a major barrier to advancing AI-based RE research~\cite{ref_dealingdata,ref_replication,b11}. This persistent scarcity, combined with common issues such as small scale, class imbalance, and narrow domain coverage~\cite{ref_dealingdata,ref_which,b11}, makes RE a relevant domain for studying SDG. While we use RE classification tasks for evaluation, the approaches themselves are domain-agnostic and applicable wherever synthetic training data is needed.

Our experimental findings show that multi-sample prompting substantially improves both diversity and downstream utility, with F1-score gains of 6 to 43.8 percentage points across tasks. PACE consistently improves lexical diversity but has task-dependent effects on utility, highlighting the risks of optimizing for diversity alone. Most notably, synthetic data can match or surpass human-authored data for specific classification tasks, with improvements of up to 15.4 percentage points in F1-score.

\textbf{Contributions.} (1) We provide an empirical evaluation of multi-sample prompting and PACE-based prompt optimization for SDG, revealing that multi-sample prompting consistently improves both diversity and utility, while PACE yields diversity gains with task-dependent utility effects. (2) We present Synthline, an FM-based configurable synthetic data generator that integrates these mechanisms, released publicly for reproducibility.\footnote{\url{https://github.com/abdelkarim-elhajjami/synthline/tree/v0.1.0}}

The paper is structured as follows: Section~\ref{sec:synthline} presents Synthline and the two diversity mechanisms. Section~\ref{sec:setup} describes the experimental setup. Section~\ref{sec:results} presents the results. Section~\ref{sec:discussion} discusses the findings, their implications, and threats to validity. Section~\ref{sec:conclusion} concludes.
\section{Synthline Overview}
\label{sec:synthline}
Synthline is a FM-based generator for synthetic data~\cite{ref_synthlinev0}. Its configurable architecture allows users to control the variability of generated data through feature selection. In this paper, we extend Synthline with two mechanisms designed to address a core challenge of LLM-based SDG: the lack of output diversity caused by post-training typicality bias~\cite{zhang2025verbalized}. The first mechanism, multi-sample prompting, exploits within-context awareness to reduce repetition without any additional infrastructure. The second mechanism, PACE-based prompt optimization, automates the search for diversity-maximizing prompts through iterative actor-critic refinement, removing the burden of manual prompt engineering. We describe the FM that structures Synthline's configuration space, then detail each mechanism and the overall architecture.
\subsection{Feature Model}
Synthline is organized around four feature groups. The \texttt{Generator} feature controls data generation through LLM configuration (model selection, temperature, top-p) and the prompting approach: \texttt{samplesPerPrompt} sets the number of samples requested per prompt, while \texttt{promptApproach} selects between \texttt{Default} template-based prompting and \texttt{PACE} automated prompt optimization. When PACE is selected, additional parameters become available: \texttt{paceIterations}, \texttt{paceActors}, and \texttt{paceCandidates}. The \texttt{Artifact} feature specifies characteristics of the generated data (requirement type, specification level, source, format, domain, and language). The \texttt{MLTask} feature defines the target classification task through a label and its description. The \texttt{Output} feature specifies the format and size of the generated dataset.
\subsection{Multi-sample Prompting}
When an LLM generates a single sample per prompt, post-training alignment drives it toward the most stereotypical output for that prompt~\cite{zhang2025verbalized}. Multi-sample prompting counteracts this by requesting multiple samples within a single prompt, leveraging within-context awareness to reduce repetition~\cite{berro2025llms}. 
In Synthline, the number of samples per prompt is configurable via \texttt{samplesPerPrompt}. This required developing two distinct prompt templates: a single-sample template, and a multi-sample template that explicitly requests multiple items and instructs the LLM to return a JSON array. Beyond diversity, this approach offers an efficiency advantage: generating $n$ samples per prompt requires $1/n$ the number of API calls compared to single-sample generation~\cite{cheng2023batch}.

\subsection{PACE-based Prompt Optimization}
\begin{figure}[ht] \centering \includegraphics[width=0.9\textwidth]{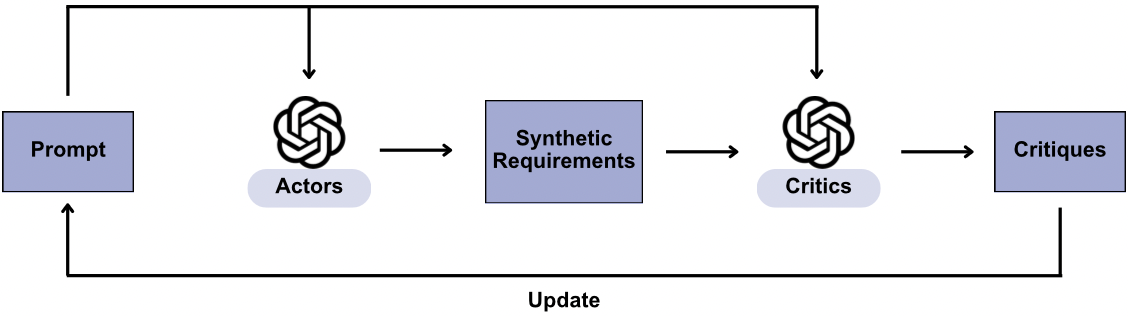} \caption{The PACE optimization process integrated into Synthline. The Actor LLM generates candidate samples from the current prompt; a scoring function evaluates the batch; the Critic LLM provides natural-language feedback; and the Actor produces a revised prompt. After the final iteration, the best-scoring prompt is forwarded to the main generation phase.} \label{fig:pace} \end{figure}

To move beyond manual prompt engineering, we integrated PACE~\cite{ref_pace} directly into the generation workflow. As illustrated in Figure~\ref{fig:pace}, PACE operates as a pre-generation optimization loop. An \textit{Actor} LLM generates a batch of samples, and a \textit{Critic} LLM evaluates them and provides feedback to iteratively refine the prompt before the main generation phase begins. More specifically, the process works as follows: (1)~the Actor receives the current prompt and generates a batch of candidate samples; (2)~each sample is scored using a configurable scoring function; (3)~the Critic receives the scored batch along with the current prompt and produces natural-language feedback identifying weaknesses and suggesting specific modifications; (4)~the Actor incorporates this feedback to generate revised prompt candidates. This cycle repeats for a fixed number of iterations, after which the highest-scoring prompt is selected for the main generation phase.

In our implementation, the scoring function was configured to maximize the average pairwise cosine distance between the Sentence-BERT embeddings~\cite{reimers2019sbert} of generated samples within a batch, thereby encouraging diversity. This purely diversity-oriented objective allows us to empirically test whether maximizing diversity at the prompt level translates into improved downstream utility.

\subsection{Architecture}
\begin{figure}[ht]
\centering
\includegraphics[width=0.9\textwidth]{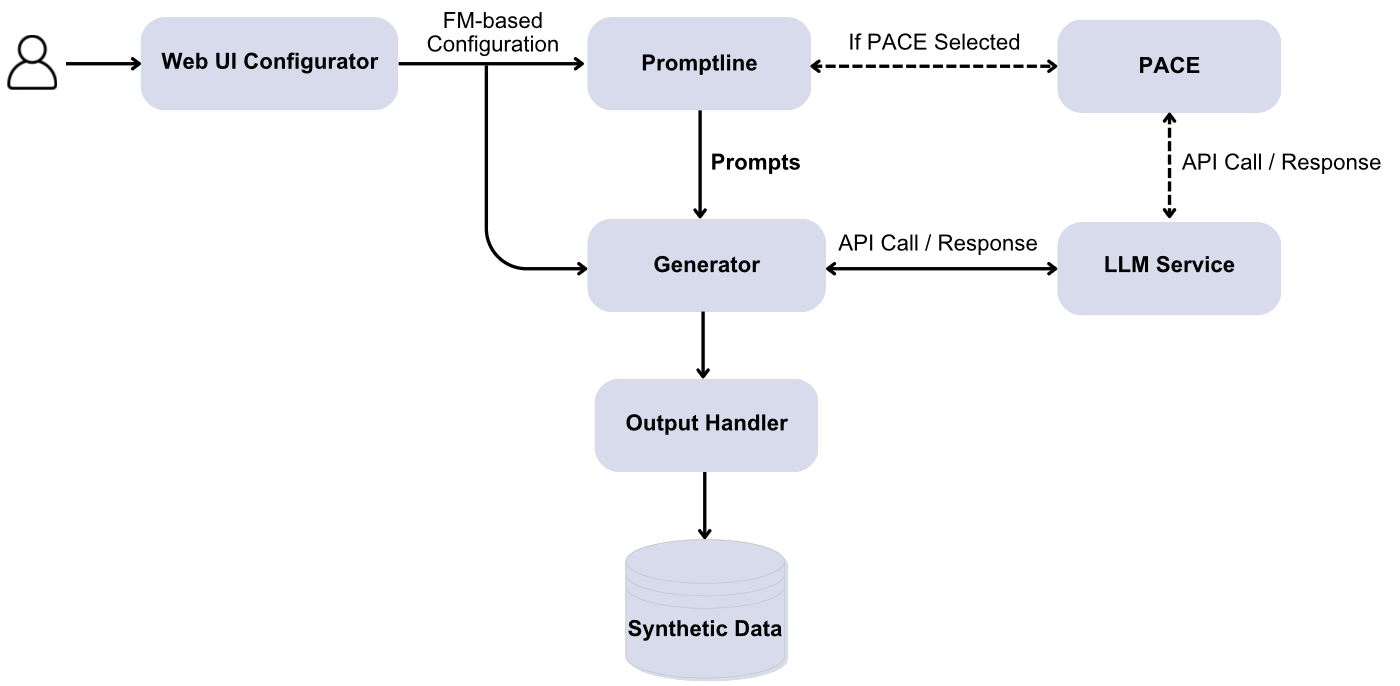}
\caption{Synthline architecture with conditional PACE optimization branch.}
\label{fig:synthlinev1}
\end{figure}

As illustrated in Figure~\ref{fig:synthlinev1}, the workflow begins with the \textit{Web UI Configurator} processing user-selected features and generating an FM-based configuration that flows to the \textit{Promptline} component. \textit{Promptline} interprets this configuration, produces all valid combinations (termed atomic configurations), then constructs corresponding prompts using single-sample or multi-sample templates based on the \texttt{samplesPerPrompt} parameter.
When PACE optimization is enabled, \textit{Promptline} invokes the \textit{PACE} module, which coordinates with the \textit{LLM Service} to perform iterative refinement before large-scale generation. When default prompting is selected, prompts bypass \textit{PACE} and flow directly to the \textit{Generator}.
Once prompts are finalized, the \textit{Generator} module orchestrates the main data generation phase, managing large-scale interaction with the \textit{LLM Service}. The requested number of samples is distributed evenly across atomic configurations, ensuring systematic coverage of the specified variability space.
\section{Experimental Setup}
\label{sec:setup}
This section describes the experimental setup used to evaluate the two diversity mechanisms presented in Section~\ref{sec:synthline}. The overall experimental design is as follows: we compare three synthetic dataset variants---single-sample generation (synthetic\_1), multi-sample generation (synthetic\_20), and PACE-optimized multi-sample generation (synthetic\_20\_pace)---against real-data baselines across four RE classification tasks. For each variant, we evaluate both the diversity of the generated data (using lexical and semantic metrics) and its downstream utility (by training BERT classifiers and measuring their performance on real test data). We first present the RE classification tasks and real-world datasets used as evaluation benchmarks, then detail the SDG configurations, and finally outline the evaluation protocol for assessing diversity and utility.
\subsection{Requirements Classification Tasks and Datasets}
This study evaluates Synthline across four distinct RE classification tasks. Each task uses established real-world datasets to provide baselines for comparison.
\subsubsection{Requirements Defects Classification}
This task involves identifying common defects in requirement specifications, crucial for early quality assurance~\cite{b41}. We adopt the six defect categories and definitions from Fazelnia et al.~\cite{b42}: Ambiguous, Directive, Non-Atomic, Non-Measurable, Optional, and Uncertain. The corresponding real dataset contains 131 samples.
\subsubsection{Functional vs. Non-Functional Classification}
This binary classification task distinguishes between functional and non-functional requirements. We use the Functional-Quality dataset from Dalpiaz et al.~\cite{ref_dalpiaz}, which contains 956 requirements from multiple open-source and closed-source projects. The dataset contains 587 functional requirements and 369 non-functional requirements.
\subsubsection{Quality vs. Non-Quality Classification}
This task focuses on identifying quality requirements. Using the same Functional-Quality dataset, a requirement is classified as \textit{Quality} when it possesses quality aspects (quality goals or quality constraints)~\cite{ref_dalpiaz}. The dataset distribution for this task includes 522 quality requirements and 434 non-quality requirements.
\subsubsection{Security vs. Non-Security Classification}
This binary classification task identifies security-related requirements. We use the SecReq dataset created by Knauss et al.~\cite{ref_secreq}, containing 510 requirements collected from three security-focused projects. The dataset contains 187 security-related requirements and 323 non-security related requirements.
\subsection{Synthetic Data Generation}
We configured Synthline to generate datasets for each of the four tasks. As the backend LLM generator, we used the OpenAI model \texttt{gpt-4.1-nano-2025-04-14}, selected for its optimal trade-off between performance, speed, and cost as of our experimentation period in late April 2025\footnote{OpenAI, Introducing GPT-4.1, \url{https://openai.com/index/gpt-4-1/}}. The base configuration used for generation is outlined in Table~\ref{tab:gen_config}.
\begin{table}[!ht]
\caption{Base Configuration Parameters for SDG.}\label{tab:gen_config}
\centering
\begin{tabular}{|l|l|}
\hline
\textbf{Feature} & \textbf{Value(s)} \\
\hline
LLM & \texttt{gpt-4.1-nano-2025-04-14} \\
Temperature & 1.0 \\
TopP & 1.0 \\
Specification Level & High-Level, Detailed \\
Requirement Source & End Users, Business Managers, Development Team, Regulatory Bodies \\
Specification Format & NL, Constrained NL, User Story \\
Language & English \\
Domain & Telecommunications, Healthcare, Enterprise Data Management \\
Output Format & CSV \\
\hline
\end{tabular}
\end{table}
To investigate our RQs, we created different dataset variants by manipulating specific generator features:
\begin{itemize}
    \item \textbf{synthetic\_1}: Generated using samplesPerPrompt = 1 and the default prompt approach.
    \item \textbf{synthetic\_20}: Generated using samplesPerPrompt = 20 and the default prompt approach.
    \item \textbf{synthetic\_20\_pace}: Generated using samplesPerPrompt = 20 and the PACE prompt approach. The PACE parameters (Table~\ref{tab:pace_params}) follow the recommendations from~\cite{ref_pace}. The same LLM (\texttt{gpt-4.1-nano-2025-04-14}) was used for both Actor and Critic roles. We configured the scoring function to maximize the average pairwise cosine distance between Sentence-BERT embeddings (using \texttt{all-MiniLM-L6-v2}) of generated samples within each batch, thereby encouraging diversity.
\end{itemize}

For each variant and classification task, 500 synthetic samples were generated per label. These samples were evenly distributed across all atomic configurations, ensuring coverage of the specified feature space.
\begin{table}[ht!]
\caption{PACE parameters used for synthetic\_20\_pace generation following~\cite{ref_pace}.}\label{tab:pace_params}
\centering
\begin{tabular}{|l|c|}
\hline
\textbf{Parameter} & \textbf{Value} \\
\hline
Number of Actor-Critic Pairs (n) & 4 \\
Number of Iterations & 3 \\
Number of Candidate Prompts per Iteration & 2 \\
Temperature (Actor, Critic \& Update) & 0 \\
Top-p (Actor, Critic \& Update) & 1 \\
Scoring Function & Cosine distance \\
\hline
\end{tabular}
\end{table}
\subsection{Evaluation Setup}
Our evaluation focuses on the quality of synthetic data as \textit{training data for ML models}. We evaluate along two complementary quality dimensions: diversity and utility. Both dimensions are necessary because they can diverge in practice: a dataset with high diversity but poor label fidelity will not produce effective classifiers, while a dataset with low diversity may overfit to narrow patterns and fail to generalize. By measuring both, we can determine whether a given mechanism genuinely improves the generated data or merely trades one quality dimension for another.
\subsubsection{Diversity Evaluation}
We calculated diversity metrics for all dataset variants (real and synthetic) for each classification task. The metrics include:
\begin{itemize}
    \item \textbf{INGF (Inter-sample N-gram Frequency):} This metric assesses lexical and phrase-level diversity by measuring how frequently the same n-grams (contiguous sequences of n words; we use trigrams, n=3) appear across different samples in the dataset. A high INGF score means the dataset contains repeated phrases across multiple samples, suggesting lower diversity at the expression level. Conversely, a lower INGF score reflects greater lexical variety and less phrase repetition between samples.
    \item \textbf{APS (Average Pairwise Similarity):} This metric captures semantic diversity by computing the average cosine similarity between Sentence-BERT embeddings (using \texttt{all-MiniLM-L6-v2}) of text pairs. A higher APS indicates that samples tend to convey semantically similar content, while a lower APS suggests richer semantic diversity.
\end{itemize}

\subsubsection{Utility Evaluation}
To evaluate utility, we adopted the standard train-on-synthetic, test-on-real paradigm. Specifically, for each synthetic dataset variant (synthetic\_1, synthetic\_20, synthetic\_20\_pace), classifiers were trained exclusively using that variant. These models, along with a baseline model trained solely on 70\% of the available real data, were then consistently evaluated on a held-out test set comprising the remaining 30\% of the real data specific to each classification task.
We used BERT-base-uncased~\cite{ref_bert} as the downstream model for all experiments.
To obtain statistically reliable performance estimates and mitigate the effects of stochasticity inherent in model training, we conducted 5 independent runs for each dataset configuration, varying the random seed for each run.
For each configuration, hyperparameter optimization was performed initially using Optuna~\cite{ref_optuna} (32 trials) on a 90\%/10\% stratified train/validation split derived from the first run's data. This process identified the optimal learning rate, batch size, weight decay, and number of training epochs by maximizing the weighted F1-score on the validation set. These best hyperparameters were then held constant across all 5 runs for that configuration, ensuring consistency while assessing performance variability.
Model performance was quantified using weighted Precision, Recall, and F1-score. We use weighted F1-score as the primary summary metric because it balances precision and recall while accounting for class imbalance through class-frequency weighting, making it appropriate for datasets with uneven label distributions. The final reported values represent the mean and standard deviation calculated across the 5 independent runs for each configuration.
\section{Results}
\label{sec:results}
This section presents the experimental results addressing our three research questions. We first examine the impact of multi-sample prompting on synthetic data quality (RQ1), then evaluate the effect of PACE-based prompt optimization (RQ2), and finally compare synthetic data against human-authored requirements for classification tasks (RQ3).

\textbf{Reading guide for tables.} In all results tables, wP, wR, and wF1 denote weighted Precision, Recall, and F1-score respectively (mean $\pm$ standard deviation across 5 runs). INGF is the Inter-sample N-gram Frequency and APS is the Average Pairwise Similarity (both defined in Section~\ref{sec:setup}). For utility metrics, lower standard deviation indicates more stable performance; for INGF and APS, lower values indicate higher diversity. \textbf{Bold} values indicate the best result for each metric within each task comparison.
\subsection{RQ1: How does multi-sample prompting impact the diversity and utility of synthetic data for classification tasks?}
We compared datasets generated with one sample per prompt (synthetic\_1) versus twenty samples per prompt (synthetic\_20).
\begin{table}[ht!]
\caption{Comparison of Utility (Weighted Metrics) and Diversity Between Single-Sample (synthetic\_1) and Multi-Sample (synthetic\_20) Datasets.}
\label{tab:rq1}
\centering
\begin{tabular}{|l|l|l|l|l|c|c|}
\hline
\textbf{Task} & \textbf{Dataset} & \textbf{wP} & \textbf{wR} & \textbf{wF1} & \textbf{INGF} & \textbf{APS} \\
\hline
\multirow{2}{*}{Security} & Synthetic\_1 & 0.649 $\pm$ 0.27 & 0.419 $\pm$ 0.07 & 0.325 $\pm$ 0.12 & 2.420 & 0.409 \\
                          & Synthetic\_20 & \textbf{0.808 $\pm$ 0.02} & \textbf{0.756 $\pm$ 0.04} & \textbf{0.763 $\pm$ 0.04} & \textbf{1.817} & \textbf{0.293} \\
\hline
\multirow{2}{*}{Quality} & Synthetic\_1 & 0.673 $\pm$ 0.17 & 0.581 $\pm$ 0.00 & 0.430 $\pm$ 0.01 & 2.194 & 0.392 \\
                         & Synthetic\_20 & \textbf{0.700 $\pm$ 0.03} & \textbf{0.650 $\pm$ 0.01} & \textbf{0.590 $\pm$ 0.03} & \textbf{1.571} & \textbf{0.281} \\
\hline
\multirow{2}{*}{Functional} & Synthetic\_1 & \textbf{0.841 $\pm$ 0.01} & 0.369 $\pm$ 0.02 & 0.365 $\pm$ 0.04 & 2.152 & 0.393 \\
                            & Synthetic\_20 & 0.819 $\pm$ 0.01 & \textbf{0.413 $\pm$ 0.07} & \textbf{0.425 $\pm$ 0.10} & \textbf{1.626} & \textbf{0.305} \\
\hline
\multirow{2}{*}{Defects} & Synthetic\_1 & 0.481 $\pm$ 0.05 & 0.415 $\pm$ 0.04 & 0.414 $\pm$ 0.04 & 2.252 & 0.372 \\
                         & Synthetic\_20 & \textbf{0.621 $\pm$ 0.04} & \textbf{0.535 $\pm$ 0.03} & \textbf{0.526 $\pm$ 0.02} & \textbf{1.619} & \textbf{0.268} \\
\hline
\end{tabular}
\end{table}
Requesting multiple samples per prompt (synthetic\_20) improves the synthetic data across both quality dimensions: utility and diversity (Table~\ref{tab:rq1}).

\textbf{Utility.} The multi-sample approach consistently outperforms single-sample generation in nearly all metrics, with F1-score improvements ranging from 6 to 43.8 percentage points across classification tasks. The largest gain appears for Security (+43.8~pp), followed by Defects (+11.2~pp) and Quality (+16~pp). The only exception is in the functional classification task where synthetic\_1 achieved higher precision, but synthetic\_20 still delivered better overall utility with higher recall and F1-scores.

\textbf{Diversity.} Synthetic\_20 datasets show substantially lower INGF values (25--28\% reduction) indicating less repetitive phrasing, and lower APS scores (22--28\% reduction) demonstrating greater semantic diversity between samples. These improvements are consistent across all four tasks.

These findings demonstrate that generating multiple samples in a single prompt leads to synthetic datasets with both higher utility for training classifiers and greater linguistic diversity.
\subsection{RQ2: How does PACE-based prompt optimization influence the diversity and utility of synthetic data for classification tasks?}
We compared standard multi-sample generation (synthetic\_20) with PACE-optimized multi-sample generation (synthetic\_20\_pace).
\begin{table}[ht!]
\caption{Comparison of Utility (Weighted Metrics) and Diversity Between Standard Multi-Sample (synthetic\_20) and PACE-Optimized (synthetic\_20\_pace) Datasets.}
\label{tab:rq2}
\centering
\begin{tabular}{|l|l|l|l|l|c|c|}
\hline
\textbf{Task} & \textbf{Dataset} & \textbf{wP} & \textbf{wR} & \textbf{wF1} & \textbf{INGF} & \textbf{APS} \\
\hline
\multirow{2}{*}{Security} & Synthetic\_20 & \textbf{0.808 $\pm$ 0.02} & \textbf{0.756 $\pm$ 0.04} & \textbf{0.763 $\pm$ 0.04} & 1.817 & 0.293 \\
                          & Synthetic\_20\_pace & 0.798 $\pm$ 0.02 & 0.713 $\pm$ 0.06 & 0.718 $\pm$ 0.06 & \textbf{1.467} & \textbf{0.290} \\
\hline
\multirow{2}{*}{Quality} & Synthetic\_20 & 0.700 $\pm$ 0.03 & 0.650 $\pm$ 0.01 & 0.590 $\pm$ 0.03 & 1.571 & 0.281 \\
                         & Synthetic\_20\_pace & \textbf{0.743 $\pm$ 0.02} & \textbf{0.668 $\pm$ 0.02} & \textbf{0.609 $\pm$ 0.04} & \textbf{1.387} & \textbf{0.267} \\
\hline
\multirow{2}{*}{Functional} & Synthetic\_20 & 0.819 $\pm$ 0.01 & 0.413 $\pm$ 0.07 & 0.425 $\pm$ 0.10 & 1.626 & \textbf{0.305} \\
                            & Synthetic\_20\_pace & \textbf{0.858 $\pm$ 0.01} & \textbf{0.723 $\pm$ 0.06} & \textbf{0.750 $\pm$ 0.05} & \textbf{1.512} & 0.309 \\
\hline
\multirow{2}{*}{Defects} & Synthetic\_20 & \textbf{0.621 $\pm$ 0.04} & \textbf{0.535 $\pm$ 0.03} & \textbf{0.526 $\pm$ 0.02} & 1.619 & \textbf{0.268} \\
                         & Synthetic\_20\_pace & 0.460 $\pm$ 0.05 & 0.465 $\pm$ 0.02 & 0.448 $\pm$ 0.03 & \textbf{1.442} & 0.276 \\
\hline
\end{tabular}
\end{table}
The impact of PACE optimization on synthetic data quality varies across classification tasks (Table~\ref{tab:rq2}). For diversity, PACE consistently improves phrase-level diversity with lower INGF values across all tasks (10--19\% reduction), indicating less repetitive phrasing. However, its effect on semantic diversity (APS) is minimal, with only slight improvements in two tasks. Regarding utility, PACE produces mixed results: it dramatically improves classification performance for Functional requirements (F1-score increase of 32.5 percentage points) and moderately improves Quality classification (1.9 percentage point gain). However, it negatively impacts Security and Defects classification performance, with F1-score decreases of 4.5 and 7.8 percentage points respectively. These findings suggest that while PACE consistently enhances lexical diversity based on its cosine distance objective, its impact on downstream classification utility is task-dependent.
\subsection{RQ3: To what extent can synthetic data replace or surpass human-authored data for classification tasks?}
We compared the performance of models trained on the best performing synthetic dataset variant for each task against models trained on the real dataset.
\begin{table}[ht!]
\caption{Comparison of Best Synthetic Datasets vs Human-Authored Requirements for Classification Tasks.}
\label{tab:rq3_utility}
\centering
\begin{tabular}{|l|l|l|l|l|}
\hline
\textbf{Task} & \textbf{Dataset} & \textbf{wP} & \textbf{wR} & \textbf{wF1} \\
\hline
\multirow{2}{*}{Security}
& Real & 0.720 $\pm$ 0.031 & 0.682 $\pm$ 0.038 & 0.685 $\pm$ 0.035 \\
& Synthetic\_20 & \textbf{0.808 $\pm$ 0.022} & \textbf{0.756 $\pm$ 0.044} & \textbf{0.763 $\pm$ 0.043} \\
\hline
\multirow{2}{*}{Quality}
& Real & 0.712 $\pm$ 0.007 & \textbf{0.688 $\pm$ 0.014} & \textbf{0.688 $\pm$ 0.015} \\
& Synthetic\_20\_pace & \textbf{0.743 $\pm$ 0.024} & 0.668 $\pm$ 0.018 & 0.609 $\pm$ 0.036 \\
\hline
\multirow{2}{*}{Functional}
& Real & \textbf{0.891 $\pm$ 0.009} & \textbf{0.830 $\pm$ 0.017} & \textbf{0.845 $\pm$ 0.015} \\
& Synthetic\_20\_pace & 0.858 $\pm$ 0.010 & 0.723 $\pm$ 0.061 & 0.750 $\pm$ 0.053 \\
\hline
\multirow{2}{*}{Defects}
& Real & 0.372 $\pm$ 0.050 & 0.410 $\pm$ 0.034 & 0.372 $\pm$ 0.034 \\
& Synthetic\_20 & \textbf{0.621 $\pm$ 0.044} & \textbf{0.535 $\pm$ 0.025} & \textbf{0.526 $\pm$ 0.022} \\
\hline
\end{tabular}
\end{table}
Synthetic data can match or surpass human-authored data in specific classification tasks (Table~\ref{tab:rq3_utility}). For Security and Defects, the best synthetic data configurations outperform human-authored requirements, with F1-score improvements of 7.8 and 15.4 percentage points respectively. The improvement is most pronounced for Defects classification, where synthetic data delivers 41.4\% better F1-score performance. For Quality requirements, synthetic data offers stronger precision (+3.1 points) but results in a weaker F1-score ($-$7.9 points). Only for Functional requirements do human-authored examples remain definitively superior, outperforming the best synthetic dataset by 9.5 percentage points in F1-score.
These results establish that synthetic data can match or exceed human-authored data for certain classification tasks, particularly when real labeled data is limited.
\section{Discussion}
\label{sec:discussion}
Our results demonstrate that multi-sample prompting consistently improves both the diversity and utility of synthetic data, that PACE-based prompt optimization yields consistent diversity gains with task-dependent effects on utility, and that synthetic data can match or exceed human-authored data for certain classification tasks. We now discuss the originality of these findings in light of existing work, their implications for research and practice, and potential limitations.

\subsection{Empirical Contributions}
This paper presents an empirical evaluation of two mechanisms for SDG.

\paragraph{Multi-sample Prompting.}
Multi-sample prompting, requesting multiple data points within a single prompt, was shown to improve diversity in paraphrase generation by Berro et al.~\cite{berro2025llms}. Our results extend this finding to SDG for text classification: multi-sample prompting reduces both lexical repetition (25--28\% lower INGF) and semantic similarity (22--28\% lower APS) while simultaneously improving downstream classifier performance by 6--43.8 percentage points in F1-score. This dual improvement in diversity and utility is notable, as increasing diversity through standard techniques such as temperature sampling often degrades label accuracy~\cite{ref_diversityaccuracy}.

Beyond diversity, multi-sample prompting offers a practical efficiency advantage. As demonstrated by Cheng et al.~\cite{cheng2023batch}, batching multiple generation requests within a single prompt reduces both token usage and API call overhead compared to issuing separate prompts for each sample. However, this efficiency comes with a trade-off in parsing robustness: multi-sample prompts require the LLM to return structured JSON arrays, which may occasionally be malformed or incomplete, introducing a small risk of data loss during parsing.

\paragraph{Automated Prompt Optimization.}
While Freise et al.~\cite{freise2025automatic} reviewed the potential of automated prompt optimization techniques for SDG, empirical evaluation was lacking. Our experiments provide this evaluation for PACE~\cite{ref_pace}, revealing a more nuanced picture than anticipated. PACE consistently improves lexical diversity across all tasks (10--19\% lower INGF), confirming that iterative prompt refinement guided by a diversity-oriented scoring function is effective at the phrase level. However, the impact on downstream utility is task-dependent: PACE substantially improves functional requirements classification (+32.5 percentage points F1) while degrading performance on security ($-$4.5 points) and defects ($-$7.8 points) tasks.

This divergence highlights a fundamental limitation of optimizing prompts solely for diversity. Our PACE scoring function maximizes the average pairwise cosine distance between generated samples, a purely diversity-oriented objective that does not account for what the downstream classifier actually needs to learn. This suggests that effective prompt optimization for SDG requires scoring functions that balance diversity with task-specific utility signals.

This limitation also raises a cost-effectiveness concern. PACE introduces additional computational overhead through its iterative actor-critic loop: each optimization cycle requires multiple LLM calls for sample generation, embedding computation, and prompt refinement before the main generation phase even begins. When this overhead yields clear utility gains (as in functional requirements), the investment is justified. However, when the result is degraded classifier performance (as in security and defects), the additional cost produces a net negative outcome.

\subsection{Implications}
These findings carry implications beyond the RE domain in which we conducted our evaluation. The diversity-utility challenge is not specific to requirements data: any domain where LLMs are used to generate training data faces the same tension between producing varied outputs and maintaining label accuracy. Multi-sample prompting addresses this challenge at the prompting level itself, without requiring specialized infrastructure. Similarly, the lesson from PACE, that diversity-only optimization is insufficient, applies broadly to any automated prompt optimization approach for SDG, regardless of the target domain.

From a practical standpoint, the finding that synthetic data can match or exceed human-authored data for certain tasks (with improvements up to 15.4 percentage points for defects classification) demonstrates a viable path for domains where labeled data is scarce, expensive, or subject to privacy constraints. The magnitude of this improvement is partly explained by two inherent advantages of synthetic generation: the ability to produce balanced class distributions and to scale dataset size on demand. The defects dataset contains only 131 imbalanced real samples, whereas our synthetic variant provides 500 balanced samples per label. We consider these advantages not as confounding variables but as practical benefits of SDG: in real-world usage, practitioners would naturally leverage the ability to generate balanced, large-scale datasets. In contrast, tasks with larger real datasets (e.g., functional and quality with 956 samples) show that real data retains an advantage when sufficient labeled examples are already available. Even in these cases, the ability to generate large, balanced, and configurable datasets on demand, at substantially lower cost than manual annotation, makes synthetic generation a compelling complement to traditional data collection.

\subsection{Threats to Validity}
\textbf{Internal validity} threats come from our experimental design choices. The selection of \texttt{gpt-4.1-nano-2025-04-14} as the sole LLM backend introduces potential model-specific biases; results may differ with other LLMs. Our PACE parameter configuration followed recommendations from the original paper~\cite{ref_pace} but was not optimized specifically for SDG tasks.

\textbf{External validity} is limited by our focus on English-language requirements and four classification tasks, which represent only a subset of possible applications. While we argue that the mechanisms are domain-agnostic, empirical validation in other domains (e.g., healthcare, legal) remains necessary. The real datasets used are relatively small (131 to 956 samples), and our evaluation using BERT-base-uncased may not generalize to other model architectures.

\textbf{Construct validity} concerns arise from our diversity metrics (INGF and APS), which capture lexical and semantic diversity respectively but may not encompass all dimensions of meaningful diversity for training data. A dataset could score well on both metrics while still lacking coverage of important edge cases or decision boundaries relevant to classification. Additionally, using cosine distance as the sole PACE scoring function may not capture all aspects of diversity relevant to downstream classification, as discussed above.

\textbf{Conclusion validity} is affected by the limited number of experimental runs (5 per configuration). While we report means and standard deviations to account for stochasticity in model training, larger-scale experiments with formal statistical significance testing would strengthen the reliability of our findings.
\section{Conclusion}
\label{sec:conclusion}
This paper presented an empirical evaluation of multi-sample prompting and PACE-based automated prompt optimization for SDG. Through experiments across four RE classification tasks, we demonstrated that multi-sample prompting consistently improves both the diversity and the downstream utility of synthetic data, with F1-score gains of 6 to 43.8 percentage points over single-sample generation. This mechanism requires no additional infrastructure and offers inference efficiency advantages through reduced API calls. PACE-based prompt optimization consistently improves lexical diversity but produces task-dependent effects on utility, highlighting that optimizing prompts solely for diversity is insufficient and that effective scoring functions must balance diversity with task-specific utility signals. Most notably, synthetic data can match or surpass human-authored data for tasks where real labeled data is limited, with improvements of up to 15.4 percentage points in F1-score.

These contributions extend beyond the RE domain used for our evaluation. The effectiveness of multi-sample prompting and the limitations of diversity-only optimization are relevant to any domain such as healthcare, legal, or financial where LLMs are used to generate training data.

Future work should explore utility-aware scoring functions for prompt optimization and extend evaluation to other domains and languages.
\begin{credits}
\subsubsection{\discintname}
The authors have no competing interests to declare that are relevant to the content of this article.
\end{credits}

\end{document}